\documentclass[12pt]{article}
\begin{document}
\vspace*{-.6in} \thispagestyle{empty}
\begin{flushright}
CALT-68-2237\\ hep-th/9908144
\end{flushright}
\baselineskip = 18pt

\vspace{.5in} {\Large
\begin{center}
TASI Lectures on Non-BPS D-Brane Systems
\end{center}}

\begin{center}
John H. Schwarz\\ \emph{California Institute of Technology,
Pasadena, CA  91125, USA}
\end{center}
\vspace{1in}

\begin{center}
\textbf{Abstract}
\end{center}
\begin{quotation}
\noindent In this set of lectures various properties of D-branes
are discussed.  After reviewing the basics, we discuss unstable
D-brane/anti-D-brane systems, a subject pioneered by Sen.
Following him, we discuss the construction of the non-BPS
D$0$-brane in type I theory.  This state is stable since it
carries a conserved ${\bf Z}_2$ charge.  The general
classification of D-brane charges using K-theory is discussed. The
results for the type I theory, and the T-dual type I$^\prime$
theory, are emphasized. Compactification of type I on a circle or
torus gives a theory with 16 supersymmetries in 9d or 8d.  In each
case the moduli space has three branches. The spectrum of non-BPS
D-branes are different for each of these branches.  We conclude by
pointing out some problems with the type I D$7$-brane and
D$8$-brane predicted by K-theory.
\end{quotation}

\vfil \centerline{\it Presented at TASI-99 and Carg\`ese-99}

\newpage

\tableofcontents

\newpage


\section{Introduction}

In the past couple of years Sen has drawn attention to various
non-BPS D-brane configurations in type II superstring theories as
well as various type II orbifolds and orientifolds
\cite{sen194,sen019,sen170,sen141,sen111}. Two review articles
have been written \cite{sen207,lerda006}. One goal is to identify
and understand various objects that are stable but not
supersymmetric. By now the BPS D-branes that preserve half the
ambient supersymmetry have been extensively studied and are quite
well understood \cite{polchinski017,polchinski050}.  So it is time
to move on and develop an understanding of other objects that
carry conserved charges and therefore should have a stable ground
state. In these lectures we will concentrate on the simplest
examples only
--- especially those that occur in the type I superstring theory
in ten dimensions.

Non-BPS D-brane configurations can be used to give a new outlook
on BPS D-branes. For example, a non-BPS system consisting of a
type II D$p$-brane and a coincident $\overline{{\rm D}p}$-brane
can give a novel characterization of a BPS D$(p-2)$-brane if the
world-volume fields have a suitable vortex-like configuration.
One can also use other field configurations to construct unstable
D$(p-1)$-branes.  Even though they are unstable, and carry no
conserved charges, such unstable D-branes are useful for certain
purposes.

The basic observation that has arisen from these studies is that a
system of coincident D-branes and anti-D-branes is characterized
by a pair of vector bundles (one for the D-branes and one for the
anti-D-branes).  However, when keeping track of conserved charges,
one wants to allow for the possibility of brane-antibrane
annihilation. Thus one is led to introduce equivalence classes of
pairs of bundles. Such classes turn out to be elements of K-theory
groups. Thus, K-theory is the appropriate mathematical framework
for classifying conserved D-brane charges
\cite{moore230,witten188}. Of course, one has to understand the
physics to do this right. For one thing, one must figure out which
K-theory group is appropriate for classifying D-brane charges for
a particular superstring theory vacuum.  Also, the mathematical
theory may only apply when certain physical conditions are
satisfied. Specifically, one must use care in analyzing D-branes
of high dimension (low co-dimension), as we will explain.

We will be considering various unstable brane configurations for
which annihilation is expected to occur.  The signal of the
instability will always be the occurrence of one or more
``tachyons'' in the world-volume theory appropriate to the D-brane
system in question.  We will always assume that the tachyon rolls
to a stable minimum, as in the usual Higgs mechanism.  The
arguments we will give are physically motivated and plausible. But
they are only qualitative, as in most cases we do not have very
good mathematical control. This will be sufficient for our
purposes in these lectures.  It should be pointed out, however,
that there are two other approaches that have been used to give
more mathematical control in certain cases.  In the first approach
(used extensively in Sen's papers) the tachyon condensation
mechanism is translated into a well-controlled change in the
coupling of a marginal operator in a conformal field theory.
Another useful technique for characterizing D-brane systems uses
the so-called ``boundary-state'' formalism
\cite{polchinski88,callan88,ishibashi89}. It has also been applied
to the study of non-BPS D-brane systems
\cite{bergman155,bergman126}. The idea, roughly, is that
open-string partition functions can be viewed as cylindrical world
sheets whose boundary is on the D-brane. Therefore the D-brane can
be characterized by an appropriate coherent state in the
closed-string Fock space.

The stable non-BPS D-branes that we will discuss occur in theories
with 16 supercharges.  In ten dimensions this is just the type I
theory.  (Heterotic theories do not have D-branes.)  In fewer than
10 dimensions there are more possibilities, however. For example,
in 8 or 9 dimensions the moduli space of quantum vacua with 16
supercharges has three disconnected components.  We will explain
this and show that the different components can have different
types of non-BPS D-branes, even though the BPS ones are the same
for all components. In discussing the type I theory compactified
on a circle, it is sometimes illuminating to consider the T-dual
description, called type I$^\prime$.  At several places in our
discussion we will comment on the type I$^\prime$ interpretation
of the results.

\section{Review of Basics}

In these lectures we will be concerned with D-branes, which are
dynamical objects on which fundamental strings can terminate. (D
stands for Dirichlet.) Such objects occur in weakly coupled type I
and type II superstring theories, but not in heterotic string
theories. The superstring theories have world-sheet supersymmetry
for both left-movers and right-movers.  Thus, in the
Ramond--Neveu--Schwarz (RNS) formulation, the world-sheet spinor
has both left and right-moving components, and hence the
space-time spectrum has a Ramond-Ramond (RR) sector. In this
section we review some basic facts about RR gauge fields and
D-branes \cite{green87,polchinski98}.

\subsection{Type II RR gauge fields}

Superstrings in the RNS formalism have world-sheet fields $X^\mu
(\sigma, \tau)$ and $\psi^\mu (\sigma,\tau)$ that are related by
world-sheet supersymmetry.  Here $X^\mu$, which describes the
embedding of the world-sheet in the spacetime, transforms as a
space-time vector and a collection of world-sheet scalars.  Its
superpartner, $\psi^\mu$, is also a spacetime vector but a
collection of world-sheet spinors.  The 2d world-sheet conformal
anomaly cancels for D$=10$ $ (\mu = 0, 1, \ldots, 9)$.  There are
a variety of ways of showing this.  One is to note that the
anomaly contribution of $X^\mu$ is D and of $\psi^\mu$ is
$\frac{1}{2} D$ for a total of $\frac{3}{2} D$.  The
diffeomorphism and supersymmetry ghosts, $b$ and $\beta$, give
contributions of $-26$ and $+11$, respectively.  Thus the sum
vanishes for D$ = 10$.

For closed strings, parameterized by a periodic coordinate $0\leq
\sigma < 2\pi$, the world-sheet equations of motion are easily
solved in a flat spacetime.  For example, $X^\mu$ satisfies the 2d
wave equation and therefore
\begin{equation}
X^\mu (\sigma, \tau) = X_L^\mu (\sigma + \tau) + X_R^\mu (\sigma - \tau).
\end{equation}
Similarly, the two components of $\psi^\mu (\sigma, \tau)$, which
satisfy a 2d Dirac equation, are just $\psi_L^\mu (\sigma + \tau)$
and $\psi_R^\mu (\sigma - \tau)$.  $X^\mu$ is required to be
periodic in $\sigma$, and thus $X_L^\mu$ and $X_R^\mu$ each have
period $2\pi$. However, $\psi_L^\mu$ and $\psi_R^\mu$ can belong
to the periodic Ramond sector (R) or the antiperiodic
Neveu--Schwarz sector (NS).

In the R sector $\psi_L^\mu$ (or $\psi_R^\mu)$ has integer Fourier
modes, including a zero mode $\psi_0^\mu$.  The canonical
anticommutation relations derived from the world-sheet action
imply that these zero modes satisfy a Dirac algebra
\begin{equation}
\{\psi_0^\mu, \psi_0^\nu\} = \eta^{\mu\nu}.
\end{equation}
Thus, aside from a factor of $\sqrt{2}, \psi_0^\mu$ can be
identified with Dirac matrices $\gamma^\mu$ in ten dimensions.
(Such matrices are $32 \times 32$.)  Representations of this
algebra give space-time spinors (fermions). Spinors in 10d can be
taken to both Majorana and Weyl at the same time.  This means that
they are real (in a Majorana representation of the Dirac matrices)
and eigenstates of $\gamma_{11} = \gamma^0 \ldots\gamma^9$.  The
eigenvalue $(\pm 1)$ determines the chirality.  Thus the minimal
spinor in 10d has 16 real components.  For a massless spinor
on-shell particle, also satisfying a Dirac equation, only eight
components would describe independent propagating modes.  The fact
that a minimal spinor has 16 real components explains why the
number of conserved supercharges in 10d must be a multiple of 16.
It is 16 for the type I theory and 32 for the type II theories. In
the IIA theory the two 16-component supercharges have opposite
chirality, whereas in the IIB theory they have the same chirality.

The closed-string particle spectrum has four sectors, since
$\psi_L^\mu$ and $\psi_R^\mu$ can each have R or NS boundary
conditions.  The sectors that give spacetime bosons are
NS$\otimes$NS and R$\otimes$R, while the ones that give spacetime
fermions are NS$\otimes$R and R$\otimes$NS.  After
Gliozzi--Scherk--Olive (GSO) projection, which requires the
world-sheet fermion number to be odd (for left-movers and
right-movers separately), the spectrum is
supersymmetric~\cite{gliozzi77}.  In particular, it contains
zero-mass and positive-mass particles, but no tachyons.

The massless bosons in the NS$\otimes$NS sector arise from
tensoring two vectors, whereas in the R$\otimes$R sector they
arise from tensoring two spinors.  Thus in the NS$\otimes$NS
sector one finds the metric tensor $g_{\mu\nu}$, an antisymmetric
tensor (two-form) $B_{\mu\nu}$, and the dilaton $\phi$.  In the
R$\otimes $R sector one obtains a bispinor $C_{ab'}$ (for the IIA
case where the chiralities are opposite) or $C_{ab}$ (for the IIB
case where the chiralities are the same) \cite{green81}.  However,
these are reducible representations, and one can write
\begin{equation}\label{IIA}
C_{ab'} \sim \sum_{n{\rm ~odd}} \gamma_{ab'}^{\mu_{1}\ldots \mu_{n}}
C_{\mu_{1}\ldots \mu_{n}}
\end{equation}
\begin{equation}\label{IIB}
C_{ab} \sim \sum_{n{\rm ~even}} \gamma_{ab}^{\mu_{1}\ldots
\mu_{n}} C_{\mu_{1}\ldots \mu_{n}}.
\end{equation}
Here $\gamma^\mu = \sqrt{2} \psi_0^\mu$ are 10d Dirac matrices and
$\gamma^{\mu_{1}\ldots \mu_{n}}$ is an antisymmetrized product
\begin{equation}
\gamma^{\mu_{1}\ldots\mu_{n}} = \gamma^{[\mu_{1}} \gamma^{\mu_{2}} \ldots
\gamma^{\mu_{n}]}.
\end{equation}
The RR gauge fields $C_{\mu_{1}\ldots\mu_{n}}$ are totally
antisymmetric in their indices, and therefore they can also be
represented as differential forms
\begin{equation}
C^{(n)} \equiv C_{\mu_{1}\ldots\mu_{n}} dx^{\mu_{1}} \wedge \ldots \wedge
dx^{\mu_{n}}.
\end{equation}

\subsection{BPS D-branes}

The defining property of D$p$-branes is that they are
$(p+1)$-dimensional dynamical hypersurfaces in 10d spacetime ---
with $p$ spatial dimensions and time --- on which fundamental
strings can end~\cite{dai89,leigh89}. As a result, their dynamics
(for small string coupling constant $g$) can be studied using
string perturbation theory, even though they themselves are
nonperturbative excitations.  It turns out that each BPS
D$p$-brane is a source for an RR gauge field $C^{(p+1)}$ with one
unit of the corresponding RR charge~\cite{polchinski017}. With
this convention, a $\overline{{\rm D}p}$-brane has RR charge $-1$.

The stable BPS D$p$-branes in 10d Minkowski spacetime are easily
deduced by noting which RR gauge fields $C^{(p+1)}$ occur in the
spectrum.  The results are
\begin{equation}
\begin{array}{ll}
{\rm Type~ I} & p = 1, 5, 9\\
{\rm Type~ IIA} & p = 0, 2, 4, 6, 8\\
{\rm Type~ IIB} & p = -1, 1, 3, 5, 7, 9.
\end{array}
\end{equation}
Of course, for high-dimension branes $(p = 7,8,9)$ there are
restrictions on whether such D-branes can actually be present. The
story for $p = 7$ is given by F theory~\cite{vafa022}, which will
not be discussed in these lectures.  The restrictions for $p =
8,9$ will be discussed later.  The case $p = - 1$ is the
D-instanton, which is point-like in the Euclideanized theory.

The RR charges appear as central charges in the supersymmetry
algebra.  As a result, the BPS condition ensuring that half of the
supersymmetries are unbroken, relates the tension of a D$p$-brane
($T_{{\rm D}p}$) to the charge.  The result turns out to
be~\cite{polchinski017}
\begin{equation}
T_{{\rm D}p} = 2\pi \frac{m_{s}^{p+1}}{g}.
\end{equation}
The conventions are that the universal Regge slope $\alpha'$ is
the square of the fundamental string length $\ell_s$ $(\alpha' =
\ell_s^2)$, and the string coupling constant is determined by the
dilaton $\phi$, which we are assuming is constant, by $g =
e^\phi$.  We also define a string mass scale
\begin{equation}
m_s = \frac{1}{2\pi\ell_s}.
\end{equation}
Usually, we set $\ell_s = 1$.

The world-volume theory of D$p$-branes is most conveniently
presented in the Green--Schwarz (GS) formalism.  The type II
D$p$-brane theories contain superspace fields $X^\mu (\sigma)$ and
$\theta^a (\sigma)$ as well as a world-volume gauge field
$A_\alpha (\sigma)$.  Here $\sigma$ represents the $p + 1$
coordinates of the world volume, the index $a$ is a spacetime
spinor label, and the index $\alpha$ is a world-volume vector
label.  The end of an open string attached to the D-brane gives a
point-charge source for the $U(1)$ gauge field $A_\alpha$.

The effective world-volume action for a D$p$-brane, $S_{{\rm
D}p}$, can be written as a sum of two terms
\begin{equation}
S_{{\rm D}p} = S_{\rm DBI} + S_{\rm WZ},
\end{equation}
where DBI denotes Dirac-Born-Infeld and WZ denotes Wess--Zumino.
Complete world-volume actions with local kappa symmetry have been
presented elsewhere~\cite{cederwall148,aganagic249,bergshoeff173}.
Here we will simply sketch some basic features. Dropping fermions
and normalization factors,
\begin{eqnarray}
S_{\rm DBI} &\sim& \int d^{p+1} \sigma \sqrt{\det (g_{\alpha\beta}
+ F_{\alpha\beta})}\nonumber \\ S_{\rm WZ} &\sim& \int
(Ce^F)_{p+1}.
\end{eqnarray}
The notation requires some explanation.  $g_{\alpha\beta}(\sigma)$
is the induced world-volume metric
\begin{equation}
g_{\alpha\beta}(\sigma) = g_{\mu\nu} (X) \partial_\alpha X^\mu
\partial_\beta X^\nu,
\end{equation}
$F$ is the $U(1)$ field strength, and $(Ce^F)_{p+1}$ means the
$(p+1)$-form part where
\begin{equation}
C = \sum_{k} C^{(k)}.
\end{equation}
The index $k$ runs over odd integers in the IIA case and even integers in the
IIB case.

$N$ parallel static D$p$-branes also give a BPS system.  The RR
charge and the tension are both increased by a factor of $N$, so
the ratio is unchanged.  It follows that there is a precise
cancellation of forces so that the system is in neutral
equilibrium for arbitrary separations.  When the branes coincide
there is an enhanced nonabelian gauge symmetry~\cite{witten135},
which is broken when they are separated.  In the case of type II
D$p$-branes the enhanced gauge symmetry is $U(N)$ for all values
of $p$.  The type I theory can be obtained from the IIB theory by
an orientifold projection described later.  This projection also
influences the gauge symmetry on the brane system.  Thus one
obtains $O(N)$ for $p = 1,9$ and $Sp(N)$ for $p = 5$.  (In writing
$O(N)$ I am not being careful about distinguishing $O(N)$,
$SO(N)$, Spin ($N$), etc. Such issues will be important later.)
The WZ term $\int[C]_{p+1} = \int C^{(p+1)}$ shows that the
D$p$-brane is a source for $C^{(p+1)}$ with one unit of RR charge.

Like all dynamical objects, D-branes are sources for the
gravitational field giving rise to spacetime curvature.  In the
famous AdS/CFT duality~\cite{maldacena200} this geometry plays a
crucial role. There one considers a large $N$ limit holding the `t
Hooft parameter $\lambda = g_{YM}^2 N$ fixed.  The string coupling
$g = g_{YM}^2$ goes to zero in this limit.  The effect of $N$
D-branes on the geometry is controlled by
\begin{equation}
N \cdot T_{{\rm D}p} \cdot G_N \sim N\cdot \frac{1}{g} \cdot g^2
\sim g N,
\end{equation}
where $G_N$ is Newton's constant in 10d.  This combination is held
fixed in the AdS/CFT story.  However, we will be concerned with a
different limit, namely $g \rightarrow 0$ with $N$ fixed.  In this
limit the effect of the D-branes on the geometry can be ignored at
leading order, and one can simply regard them as hypersurfaces
embedded in Minkowski spacetime, which is what we will do.  This
reasoning is reliable for branes of sufficiently low dimension (or
high co-dimension) so that gravitational fields fall with distance
from the branes. Otherwise, as in the case of D$8$-branes in the
type I$^\prime$ theory, the effect on the geometry cannot be
ignored. It is also worth noting that this decoupling of the
geometry would not work for NS 5-branes, which have $T \sim
1/g^2$.

\subsection{T duality for D-branes}

Let us consider how T duality~\cite{giveon139} works for D-branes
in the simplest setting, namely when the spacetime is $R^9 \times
S^1$, with a single circular dimension $x^9 \equiv x$ of radius
$R$. The corresponding string world-sheet field $X(\sigma,\tau)$
then takes the form
\begin{equation}
X = mR\sigma + \frac{n}{R} \tau + ~{\rm periodic~terms}.
\end{equation}
This describes a closed string with winding number $=m$ and
Kaluza--Klein (KK) momentum $= n$.  Decomposing $X$ into
left-moving and right-moving parts $X_L (\sigma +\tau)$ and $X_R
(\sigma - \tau)$,
\begin{eqnarray}
X_L &=& \frac{1}{2} \left(mR + \frac{n}{R}\right) (\sigma +\tau) +\ldots
\nonumber \\
X_R &=& \frac{1}{2} \left(mR - \frac{n}{R}\right) (\sigma -\tau) +\ldots .
\end{eqnarray}
Under a T-duality transformation $X_R \rightarrow - X_R$, while
$X_L$ is unchanged.  Thus
\begin{equation}
X = X_L + X_R \rightarrow X_L - X_R = \frac{n}{R} \sigma + m R\tau + \ldots.
\end{equation}
Comparing with the original $X$, we see that this describes a
closed string on a circle of radius $1/R$ with winding number $=
n$ and KK momentum $= m$.

World-sheet supersymmetry requires that the transformation $X_R
\rightarrow - X_R$ be accompanied by $\psi_R^9 \rightarrow -
\psi_R^9$.  This has the consequence of reversing the right-moving
chirality projection, in particular $\gamma_{11} \rightarrow -
\gamma_{11}$.  It therefore follows that IIA $\leftrightarrow$
IIB. Thus, to recapitulate, the IIA theory compactified on a
circle of radius $R$ is equivalent to the IIB theory on a circle
of radius $1/R$ and the role of winding number and KK excitation
number is interchanged under this map.

Now let us examine the implications for D-branes.  Suppose, for
example, we start with a pair of parallel D$p$-branes with
positions given by
\begin{equation}
\# 1:  X^i = d_1^i \quad i = p + 1, \ldots, 9
\end{equation}
\begin{equation}
\# 2:  X^i = d_2^i \quad i =p + 1, \ldots, 9.
\end{equation}
Thus the branes fill the dimensions $X^m$, $m=0, 1, \ldots, p$ and
are localized in the other $9 - p$ dimensions.  An open string
connecting the two D-branes satisfies Neumann boundary conditions
in the first $p+1$ dimensions
\begin{equation}
\partial_\sigma X^m|_{\sigma = \pi} = 0,\quad m = 0, \ldots, p
\end{equation}
and Dirichlet boundary conditions in the remaining $9 - p$ dimensions
\begin{equation}
X^i|_{\sigma = 0} = d_1^i, \quad X^i|_{\sigma = \pi} = d_2^i,\quad
i = p + 1, \ldots, 9.
\end{equation}
Solving the 2d wave equation with these boundary conditions gives
\begin{equation}
X^m = x^m + p^m \tau + i \sum_{n\not= 0} \frac{\alpha_n^m}{n} \cos
n \sigma \, e^{-in\tau}
\end{equation}
\begin{equation}
X^i = d_1^i + (d_2^i - d_1^i) \frac{\sigma}{\pi} + \sum_{n\not=0}
\frac{\alpha_n^i}{n} \sin n \sigma \, e^{-in\tau}.
\end{equation}

Suppose now that $X^9$ is a circle as before.  What happens to the
D$p$-branes under the T-duality transformation?  It is easy to see
that $X_R \rightarrow - X_R$ simply interchanges the Dirichlet and
Neumann boundary conditions. Therefore the D-branes, which we
assumed were localized on the original circle of radius $R$, are
wrapped on the T-dual circle of radius $1/R$.  Thus in the T-dual
theory there are D$(p+1)$-branes, wrapped on the circular
dimension. Thus the general rule is that under T duality:
unwrapped D$p$ $\leftrightarrow$ wrapped D$(p+1)$. This meshes
nicely with the fact that BPS D$p$-branes exist in the IIA theory
for even values of $p$ and in the IIB theory for odd values of
$p$.

Now a question arises.  In the T-dual description, where the
D-branes are wrapped on the dual circle, how are the positions on
the original circle encoded?  Clearly, this information must not
be lost, since the transformation should be invertible.  The key
to answering this question is to recall that T duality
interchanges winding number and KK excitation number.  However,
the original configuration has charged states with fractional
winding associated to open strings that connect the two D-branes.
So in the dual description these charge states should have
fractional KK number.   This is achieved by Wilson lines $\exp [i
\oint A_9]$ in the world-volume gauge theories of the wrapped
D-branes.  In other words, the component of the gauge fields along
the circle encodes the position of the original D$p$-branes on the
circle.

\subsection{Type I superstrings}

The modern construction of type I superstring theory (not the way
it was originally developed) is as an ``orientifold projection''
of the type IIB theory~\cite{sagnotti87,pradisi89,horava89}.
Recall that in the type IIB theory, the fermions associated with
left-movers and right-movers have the same chirality.  As a result
the world-sheet theory has a ${\bf Z}_2$ symmetry corresponding to
interchange of all left-movers and right-movers
\begin{equation}
X_L^\mu \leftrightarrow X_R^\mu,\quad \psi_L^\mu \leftrightarrow
\psi_R^\mu.
\end{equation}
Equivalently, one can speak of world-sheet parity $\Omega$, which
reflects the spatial world-sheet coordinate $(\sigma \rightarrow -
\sigma$ or $2\pi -\sigma)$.  One can, therefore, project the
spectrum of the theory onto a left-right symmetric subspace by
using the projection operator ${1 \over 2} (1 + \Omega)$.  The
result of this is to give a closed string with a restricted set of
excitations.  This is the type I closed-string, which is
unoriented.

The interacting theory of type I closed strings is not consistent
by itself. This fact can be understood in a variety of ways.  For
one thing it is a chiral theory with $N = 1$ supersymmetry (16
supercharges), and it has gravitational anomalies.  To achieve
consistency one must add a ``twisted sector,'' which consists of
open strings.  These are strings whose ends are associated to the
fixed points of the orientifold projection ($\sigma = 0$ and
$\sigma = \pi)$. They are also unoriented.  It is known from the
anomaly cancellation requirement that the open strings should
carry $SO(32)$ Chan--Paton charges at their ends, since this is
the only Chan--Paton gauge group for which the anomalies can be
cancelled~\cite{green84}. ($E_8 \times E_8$ cannot be realized in
this way.)

The orientifold projection construction can be interpreted as
giving rise to a spacetime filling orientifold plane --- an ${\rm
O}9$-plane.  The reason it fills the entire space-time is that
every $x^\mu$ is a fixed point of $\Omega$, which does not act on
$x^\mu$.  This ${\rm O}9$-plane carries $-32$ units of RR charge.
This charge needs to be cancelled.  This is achieved by adding 32
D$9$-branes, which also fill the entire spacetime.  The $SO(32)$
gauge fields arise as zero modes of open strings that connect any
pair of D-branes.  (These are the same open strings that were
introduced earlier.)  In type II theories coincident D-branes give
$U(N)$ gauge groups, because the open strings are oriented. In the
type I theory they are unoriented, and therefore one obtains
orthogonal or symplectic gauge groups.

\subsection{T-dual description of the type I theory}

In Section 2.3 we saw that T duality on a circle $S^1$ corresponds
to $X_R \rightarrow - X_R$ (and $\psi_R \rightarrow - \psi_R)$,
and this implies that
\begin{equation}
X = X_L + X_R \rightarrow X' = X_L - X_R.
\end{equation}
In the case of type II theories, $X'$ describes a dual circle
$\tilde{S}^1$ of radius $R'= 1/R$. The type I theory is
constructed by gauging the world-sheet parity symmetry $\Omega$ of
the type IIB theory.  This corresponds to $X_L \leftrightarrow
X_R$ (and $\psi_L \leftrightarrow \psi_R$).  Thus, we see that in
the T-dual description the action of $\Omega$ gives $X'
\rightarrow - X'$.  This gauging therefore gives an orbifold
projection of the dual circle: $\tilde{S}^1/{\bf
Z}_2$~\cite{polchinski169}. This orbifold describes an interval
\begin{equation}
0 \leq X' \leq \pi R'.
\end{equation}
An alternative viewpoint is that the entire circle $\tilde{S}^1$
is present, but that what happens on the semicircle $\pi R' \leq
X' \leq 2\pi R'$ is determined by the other semicircle.  More
generally, T duality along an $n$-torus $T^n$ results in an
orbifold $\tilde{T}^n/{\mathcal I}_n$.  Here ${\mathcal I}_n$
denotes the simultaneous inversion of all $n$ coordinates of the
torus $\tilde{T}^n$.  More precisely, this inversion is carried
out simultaneously with the orientation reversal $\Omega$ of the
non-compact coordinates $(X_L^\mu \leftrightarrow X_R^\mu)$.  The
fermions are transformed at the same time.

In the case of a circle $(n = 1)$, the T-dual theory can be regarded as type
IIA orientifold of the form
\begin{equation}
(R^9 \times \tilde{S}^1)/\Omega \cdot {\mathcal I}_1.
\end{equation}
The combined operation $\Omega \cdot {\mathcal I}_1$ is the
relevant ${\bf Z}_2$ projection in this case.  The resulting
theory
--- the T-dual description of type I theory compactified on a
circle
--- is called the type I$^\prime$ theory.  (Sometimes the
alternative name type IA is used.)  T duality in this setting is
the equivalence of the type I$^\prime$ theory with the type IIB
orientifold
\begin{equation}
(R^9 \times S^1)/\Omega,
\end{equation}
which is the compactified type I theory.

The ends of the interval $X' = 0, \pi R'$ are the fixed-point set
of the IIA orientifold projection, and, therefore, they are the
locations of orientifold planes  --- ${\rm O}8$-planes.  Each of
these orientifold planes can be regarded as carrying $-8$ units of
RR charge.  Consistency of the type I$^\prime$ theory then
requires adding 16 D$8$-branes to cancel the RR charges.  They are
parallel to the ${\rm O}8$-planes, which means that they are
localized at points in the interval $0 \leq X' \leq \pi R'$, and
fill the other nine dimensions.  In addition there are 16 ``mirror
image'' D$8$-branes on the other half of the circle ($\pi R'\leq
X'\leq 2\pi R'$). Altogether, these 32 D$8$-branes are the T-duals
of the 32 D$9$-branes in the type I description.

The position of the D$8$-branes along the interval in the type
I$^\prime$ description are determined in the type I description by
Wilson lines.  The Wilson line is an element of $SO(32)$ and by an
equivalence transformation can be brought to a canonical form in
which it is written as 16 blocks of $SO(2)$ matrices, which are
characterized by angles
\begin{equation}
\theta_i = X'_i/R', \quad i = 1,2, \ldots , 16.
\end{equation}
The $X'_i$ are then the positions of the 16 dual D$8$-branes, and
their mirror images are located at $2\pi R' - X'_i$.

By introducing a Wilson line into the compactified type I theory,
one breaks the $SO(32)$ gauge group to the subgroup that commutes
with the Wilson line matrix.  Expressed in terms of the type
I${}^\prime$ description, this gives the following rules:
\begin{itemize}
\item When $n$ D$8$-branes coincide in the interior of the interval, they give
an unbroken $U(n)$ gauge group.

\item When $n$ D$8$-branes coincide with one of the ${\rm O}8$-planes, they give an
unbroken $SO(2n)$ gauge group.
\end{itemize}
In both cases the gauge bosons arise as zero modes of D$8-{\rm
D}8$ open strings. In the second case, the additional symmetry
enhancement can be traced to open strings connecting D$8$-branes
to mirror-image D$8$-branes.

The case of a trivial Wilson line $(W = \pm 1)$ corresponds to
having all 16 D$8$-branes coincide with one of the ${\rm
O}8$-planes, which gives the expected unbroken $SO(32)$ gauge
symmetry.  It should be noted that there are two additional
$U(1)$'s that cannot be pictured in terms of D-branes.  Rather
they are associated to gauge fields $g_{\mu 9}$ and $B_{\mu 9}$
that are defined in the bulk.  One linear combination of these
gauge fields belongs to the 9d supergravity multiplet.  Another
linear combination belongs to a 9d vector supermultiplet.  The
latter can participate in further symmetry enhancement in special
cases~\cite{bergman098,bachas086}. For example, the Wilson line
\begin{equation}
W = \left(\begin{array}{cc}
I_{16 + 2N} & 0\\
0 & -I_{16-2N}\end{array}\right)
\end{equation}
generically gives the gauge symmetry
\begin{equation}
SO(16 + 2 N) \times SO(16 - 2N) \times U(1)^2.
\end{equation}
However, from the S-dual perturbative heterotic theory one knows that there is
additional gauge symmetry enhancement.
\begin{equation}
SO(16 - 2N) \times U(1) \rightarrow E_{9 - N},
\end{equation}
for a specific value of the compactification radius.  In type I
units, this radius is given by $R^2 = gN/8$.  I will not explore
this issue further in these lectures.

\section{D-Brane Anti-D-Brane Systems}

When one superposes a BPS D$p$-brane and a BPS $\overline{{\rm
D}p}$-brane the system is no longer BPS, and it is unstable.  The
basic idea is that the open string connecting them has a spectrum
based on a GSO projection that is the reverse of the usual one,
and as a result its ground state is a tachyon, which signals an
instability.  One method of deriving this result is to consider
starting with a pair of coincident D$p$-branes and then imagine
rotating one of them by an angle $\pi$ thereby turning it into an
$\overline{{\rm D}p}$-brane~\cite{dealwis080}.  At intermediate
angles $\theta$ the intersection takes place on $p-1$ spatial
dimensions. By working through the implications of Dirichlet and
Neumann boundary conditions for arbitrary angles $\theta$ and
continuing from $\theta = 0$ to $\theta = \pi$, one can deduce the
reversal of the GSO projection.  The occurrence of a tachyon at
$\theta = \pi$ is physically sensible, of course, as it signals
the possibility of annihilation, which is no longer prevented by
charge conservation.

We will begin by discussing such systems for type II theories and
then move on to the type I theory.  In each case, we will be
interested in exploring the possibility of setting up field
configurations on the D$p + \overline{{\rm D}p}$ system such that
something stable survives after the annihilation takes place.

\subsection{Type II D$p + \overline{{\rm D}p}$ systems}

In the case of type II theories the open strings connecting
D-branes are oriented, and as a result the tachyonic mode of the
D$p - \overline{{\rm D}p}$ open string is complex.  Denoting the
tachyon field in the $(p + 1)$-dimensional world volume by $T$,
one can combine it with the gauge fields in a ``superconnection''
of the form
\begin{equation}
{\mathcal A} = \left(\begin{array}{cc} A & T\\ \bar T & A'
\end{array} \right).
\end{equation}
Here $A$ is the $U(1)$ gauge field associated to the D$p$-brane
and $A'$ is the $U(1)$ gauge field associated to the
$\overline{{\rm D}p}$-brane. The tachyon is charged with respect
to both of them.  More generally, one could consider $N$
D$p$-branes and $N'$ $ \overline{{\rm D}p}$-branes.  Then $A$
would be $U(N)$ gauge fields, $A'$ would be $U(N')$ gauge fields,
and $T$ would be an $N\times N'$ matrix transforming as $(N, \bar
N')$. To start with, we will restrict consideration to the case $N
= N' = 1$.

As an explicit example of what can happen, let us (following
Sen~\cite{sen170}) consider a D$2 + \overline{{\rm D}2}$ system
wrapped on a rectangular torus with radii $R_1, R_2$.  As we
discussed earlier, the WZ term of the D$2$-brane is
\begin{equation}
\int (Ce^F)_3 = \int_{T^2 \times R} (C_3 + C_1 \wedge F).
\end{equation}
This formula shows that magnetic flux through the torus ---
$\int_{T^2} F$ --- is a source of $C_1$.  In other words, a
D$2$-brane with magnetic flux carries D$0$-brane charge.

Suppose that the D$2$ and the $\overline{{\rm D}2}$ each have one
unit of D$0$-brane charge.  Assuming constant fields, this means
that
\begin{equation}
F_{12} = F'_{12} = \frac{2\pi}{V},
\end{equation}
where $V = (2\pi R_1) (2\pi R_2)$ is the volume of the torus. Let
us now examine the system described above from a T-dual viewpoint.
In the T-dual description (which is also type IIA, since the torus
has even dimension) we get a dual torus $\tilde{T}^2$ with radii
$\tilde{R}_i = 1/R_i$.  By matching coupling constants in eight
dimensions one learns that
\begin{equation}
\frac{V}{g^2} = \frac{\tilde{V}}{\tilde{g}^2},
\end{equation}
and hence that
\begin{equation}
\tilde{g} = g/(R_1 R_2).
\end{equation}
Since the T-duality transformation interchanges D$0
\leftrightarrow {\rm D}2$, the fluxes that gave D$0$-brane charges
imply that the dual system has two D$2$-branes wrapping
$\tilde{T}_2$. Being two coincident type II D-branes, the system
has a $U(2)$ gauge symmetry.  The original D$2$ and
$\overline{{\rm D}2}$ give rise to a D$0$ on one D$2$ and a
$\overline{{\rm D}0}$ on the other. Thus the dual magnetic flux is
\begin{equation}\label{flux}
\tilde{F}_{12} = \frac{2\pi}{\tilde{V}} \left(\begin{array}{cc} 1 & 0\\ 0 & -1
\end{array} \right).
\end{equation}
The original D$2 + \overline{{\rm D}2}$ system had a complex
tachyon, signaling instability, and the T-duality transformation
certainly doesn't change that fact.  However, now the instability
is attributable to structure of the flux matrix.  The minimum
energy configuration corresponds to $\tilde{F}_{12} = 0$, which
corresponds to annihilation of the D$0 + \overline{{\rm D}0}$.
The nontrivial fact, which Sen has demonstrated, is that the flux
in eq. (\ref{flux}) is continuously connected to $\tilde{F}_{12} =
0$. So this deformation corresponds to the tachyon rolling to a
minimum in the dual picture.  We are thus left with a pair of
wrapped D$2$-branes and no flux.

The total mass at the minimum (ignoring the contribution of radiation emitted
in the annihilation process) is
\begin{equation}
M = 2 \tilde{V} \tilde{T}_{{\rm D}2} = 2 \cdot 4\pi^2 R_1 R_2
\cdot \frac{1}{4\pi^2\tilde{g}} = \frac{2}{g} =2 T_{{\rm D}0}.
\end{equation}
This calculation shows that the mass is just that of two
D$0$-branes in the original picture.  This means that on the
original torus the mass density $M/V$ vanishes in the limit of
large $V$.  This implies that the tachyon potential at its minimum
precisely cancels the D-brane tension.  Specifically
\begin{equation}
2 T_{{\rm D}2} + V(T_0) = 0,
\end{equation}
where $|T| = T_0$ is the (assumed) position of the minimum. This
reasoning generalizes to arbitrary dimensions of the branes. So
the general conclusion is that a coincident D$p$-brane $+
\overline{{\rm D}p}$-brane system with the tachyon field at a
minimum is equivalent to pure vacuum.  This result makes good
sense.
 It is also a very insightful viewpoint.

\subsection{Vortex solutions}

Let us now consider the D$2 + \overline{{\rm D}2}$ system in ${\bf
R}^{10}$. As we already noted, the tachyon $T$ has charge $(1,
-1)$ with respect to the $U(1) \times U(1)$ gauge symmetry.  The
tachyon potential $V(T)$ is gauge invariant. Therefore, it can
only depend on the magnitude $|T|$, and its minimum is assumed to
occur for $|T| = T_0$.  Its kinetic term is $|D_\mu T|^2$, where
\begin{equation}
D_\mu T = (\partial_\mu - i A_\mu + i A'_\mu) T.
\end{equation}
A single D0-brane can be described as a vortex solution on the
infinite D$2 + \overline{{\rm D}2}$ system~\cite{sen141}.  Working
in polar coordinates $(r, \theta)$ we want to describe a vortex
localized in the vicinity of $r = 0$.  We want the field
configuration to describe pure vacuum for large $r$, but in a
topologically nontrivial way.  This is achieved by requiring that
as $r \rightarrow \infty$
\[ T \sim T_0 e^{i\theta}\]
\begin{equation}
A_\theta - A_\theta^{\prime} \sim 1.
\end{equation}
These ensure that $D_\mu T \rightarrow 0$ and $V(T) \rightarrow
V(T_0)$, so that the energy density cancels for large $r$ leaving
a soliton in the small $r$ region.  We can see that this
configuration has the quantum numbers of a D$0$-brane, since
\begin{equation}
\oint_{\rm large~circle} (A_\theta - A'_\theta) d\theta = 2\pi
\end{equation}
implies that
\begin{equation}
\int (F_{12} - F'_{12}) dx^1 dx^2 = 2\pi.
\end{equation}
As we see it is only the relative $U(1)$ group that matters.  We
could, for example, set $A' = 0$.  In any case, we conclude that
there is one unit of magnetic flux, as required for a D$0$-brane.

This result extends trivially to larger values of $p$ by simply
adding additional inert directions.  The conclusion is that a
D$(p-2)$-brane can be described as a vortex solution in a D$p +
\overline{{\rm D}p}$ system.  The fact that it carries the correct
RR charge can be understood as a consequence of the coupling
\begin{equation}\label{Tcoupling}
\int C_{p-1}~ _\wedge dT_\wedge d \bar T,
\end{equation}
which is required for consistency of the picture.  This term can
be elegantly described using the superconnection ${\mathcal A}$,
following Kennedy and Wilkins~\cite{kennedy195}.  First one
introduces the supercurvature
\begin{equation}
{\mathcal F} = d {\mathcal A} + {\mathcal A}_\wedge {\mathcal A} =
\left(\begin{array}{cc} F - T\bar T & DT \cr \overline{DT} & F' -
\bar T T\end{array} \right),
\end{equation}
which has even-dimensional forms on the diagonal, and
odd-dimensional ones on the off-diagonal.  Using this one can
construct a suggestive generalization of the WZ term to
brane-antibrane systems, namely
\begin{equation}
S_{\rm WZ} = \int C_\wedge STr (e^{{\mathcal F}}).
\end{equation}
The symbol $STr$ is the usual super-trace for graded algebras.  This contains
the usual terms
\begin{equation}
\int C_\wedge (e^F - e^{F'}),
\end{equation}
where the minus sign reflects the opposite RR charges.  It also
contains the term in eq.~(\ref{Tcoupling}), as well as some
others. It would be interesting to see whether this nice formula
reflects some new symmetry.

\subsection{Unstable type II D-branes}

We have seen that a suitably chosen vortex in a coincident D$p +
\overline{{\rm D}p}$ system describes a stable D$(p-2)$-brane. Now
(following Sen~\cite{sen141}) we would like to consider a
different field configuration that describes an unstable soliton.
Recall that at its minimum $(|T| = T_0)$ the energy density
associated with the tachyon potential exactly cancels the tension
of the branes, $V(T_0) + 2T_{{\rm D}p} = 0$, leaving a
configuration equivalent to pure vacuum.  The proposal is to
construct a D$(p-1)$-brane as a domain-wall on the D$p +
\overline{{\rm D}p}$ system.
 This means that if $x$ denotes a Cartesian coordinate, we want to construct a
soliton localized in the vicinity of $x = 0$.

The way to construct a localized domain wall is by choosing a kink
configuration for the tachyon field.  Specifically consider
choosing ${\rm Im}\, T = 0$ and ${\rm Re}\, T = T_0 \tanh (x/a)$.
This has the property that $T \rightarrow T_0$ for $x \rightarrow
+ \infty$ and $T \rightarrow - T_0$ for $x \rightarrow - \infty$.
Also, $|T|$ differs from $T_0$ appreciably only in a region of
thickness $a$ in the vicinity of $x = 0$.  The precise functional
form of $T(x)$ is not important.  This configuration gives pure
vacuum except in the vicinity of $x = 0$, so it can be interpreted
as a D$(p-1)$-brane.  This domain wall soliton is unstable because
the field configuration we have chosen is topologically trivial.
To understand this, we note that the vacuum manifold given by $|T|
= T_0$ is a circle $(S^1)$.  However, a circle is connected
$(\pi^0 (S^1)$ is trivial).  Therefore, by turning on ${\rm Im}\,
T$ and increasing the thickness $a$ this configuration can be
continuously deformed to pure vacuum. Thus we conclude that this
D-brane is unstable. Since it carries no conserved charge it can
decay into neutral radiation (such as gravitons).  Despite their
instability, Sen has demonstrated that it is useful to consider
these objects. They can be defined in type II theories for all
``wrong'' values of $p$ (odd for IIA and even for IIB).

The instability of the D$(p-1)$-brane is reflected in the fact
that its world-volume theory contains a real tachyon.  The reason
there is one real tachyon is that there is one direction of
instability for unwinding the kink in the vacuum moduli space.  It
can also be understood as the ground state of the open string
connecting the D-brane to itself.  In any case, we can use this
tachyon to repeat the kink construction.  This gives us a domain
wall on the unstable D$(p-1)$-brane, which is interpreted as a
D$(p-2)$-brane.  This kink is topologically stable, however.  The
tachyon on the D$(p-1)$-brane is real and the vacuum moduli space
in this case is two disconnected points $T = \pm T_0$. So this
construction gives a stable D$(p-2)$-brane, which carries a
conserved charge.

We already have a candidate for what this D$(p-2)$-brane is.  It
is exactly the same stable BPS D-brane that we constructed as a
vortex solution in Section 3.2.  Sen has shown that this two-step
kink construction is equivalent to the vortex solution.

\subsection{Type I D$p + \overline{{\rm D}p}$ systems}

The type I theory has three types of BPS D$p$-branes, namely $p =
1,5,9$.  Just as we constructed the type II D$(p-2)$-brane as a
soliton in a D$p + \overline{{\rm D}p}$ system, so we can
construct type I D-branes as solitons in higher-dimensional
brane-antibrane systems.  In this section we will sketch the
construction of the D$1$-brane as an instanton-like configuration
in a D$5 + \overline{{\rm D}5}$ system~\cite{sen141} and the
construction of the D$5$-brane as an instanton-like configuration
in a  D$9 + \overline{{\rm D}9}$ system~\cite{witten188}. The two
constructions are not completely identical (unlike for the type II
theories), and therefore need to be considered separately.

Before proceeding, let us recall a few basic facts about the type
I D-branes. First of all, the D$1$-brane or D-string is nothing
but the $SO(32)$ heterotic string in disguise.  Recall that the
$SO(32)$ heterotic theory is S-dual to the type I
theory~\cite{witten124,polchinski169}, which means that the
coupling constants are related by $g_h = 1/g$.  In talking about
the type I theory it is implicit that $g$ is small, so that we can
use perturbative analysis.  In type I string units, the type I
fundamental string has tension $1/2 \pi$, whereas the D-string has
tension $1/(2\pi g)$ and thus appears as a nonperturbative object.
Even so, the D-string can be identified as the heterotic string
continued to strong coupling. This string is BPS, carrying a
conserved charge.  This charge is an RR charge from the type I
viewpoint, though it is NS from the heterotic viewpoint.  The type
I D$5$-brane is the electromagnetic dual of the D-string and is
also BPS. We also need to know the gauge groups that are
associated to the D-brane world volumes. It turns out that $N$
coincident D-strings give $O(N)$ gauge symmetry, whereas $N$
coincident D$5$-branes gives $Sp(N) \equiv USp (2N)$ gauge
symmetry~\cite{witten030}.  These facts can be derived by
considering coincident type IIB D-branes and applying the
appropriate orientifold projection.

Let us now consider a coincident type I D$5 + \overline{{\rm D}5}$
system.  The world-volume theory has $SU(2) \times SU(2)$ gauge
symmetry since each brane has an $Sp(1) = SU(2)$ gauge field on
its world volume.  The ground state of the open string connecting
the D$5$ to the $\overline{{\rm D}5}$ gives a tachyon $T$ with
$(2,2)$ quantum numbers.  It can be parameterized in the form
\begin{equation}
T = \lambda W,
\end{equation}
where $\lambda$ is a positive real number and $W$ is an $SU(2)$
matrix.  The two $SU(2)$'s act on $T$ by left and right
multiplication. The tachyon potential $V(T)$ is required to be
gauge invariant, which implies that it is a function of $\lambda$
only.  As in the corresponding type II constructions, we assume
that it has an isolated minimum at some value $\lambda =
\lambda_0$ and that if $T = \lambda_0 W$, with $W$ fixed, and all
gauge fields vanishing, then the D$5 + \overline{{\rm D}5}$ system
is equivalent to pure vacuum. As before, this requires that
\begin{equation}
V(\lambda_0) + 2 T_{{\rm D}5} = 0,
\end{equation}
so that the total energy density cancels at the minimum.

We now wish to construct a field configuration on the D$5 +
\overline{{\rm D}5}$ system that describes a string-like soliton,
which can be identified as the D-string.  Thus we will assume
symmetry in one Cartesian direction.  The four orthogonal spatial
directions can be described by a distance $r$ from the core of the
string and a sphere $S^3$.  So we want the soliton's energy to be
concentrated in the vicinity of $r = 0$ and to approach zero as $r
\rightarrow \infty$.  At infinity, we certainly need that $\lambda
\rightarrow \lambda_0$, but we also need that the tachyon kinetic
energy
\begin{equation}
Tr [(D_\mu T)^\dagger D^\mu T],
\end{equation}
should vanish as well.  Here
\begin{equation}
D_\mu T = \partial_\mu T + i (A_\mu T - T A'_\mu),
\end{equation}
and $A_\mu$ is the $SU(2)$ gauge field of the D$5$-brane and
$A'_\mu$ is the $SU(2)$ gauge field of the $\overline{{\rm
D}5}$-brane. The story is very similar to the type II vortex
solution, except that it is now based on an instanton-like
construction.

We require that for large $r$
\begin{equation}
T \sim \lambda_0 U,
\end{equation}
where the $SU(2)$ element $U$ is identified with the spatial
$S^3$. Then the tachyon kinetic energy can be made to vanish at
the same time by requiring that
\begin{eqnarray}
A_\mu &\sim & i \partial_\mu U \cdot U^{-1}\nonumber \\ A'_\mu
&\sim & 0.
\end{eqnarray}
The first formula is the standard instanton configuration and gives one unit of
instanton number
\begin{equation}
\frac{1}{8\pi} \int_{R^{4}} tr (F_\wedge F) = 1.
\end{equation}
Then the structure of the WZ term implies that this configuration
carries one unit of D$1$-brane RR charge and can be identified as
the D-string.  The various properties of the D-string, such as it
various zero modes, can also be derived from this construction,
but we will not pursue that here.

The construction described above is based on the work of Sen.
However, it should be pointed out that something closely related
was done earlier by Douglas~\cite{douglas198}.  Specifically,
Douglas considered an instanton configuration on a single
D$5$-brane and argued that this describes a D-string bound to the
D$5$-brane.

Let us now turn to the construction of the D$5$-brane as a
soliton.  For this purpose we need to consider four D$9$-branes
and four $\overline{{\rm D}9}$-branes in addition to the usual 32
D$9$-branes that are present in the type I vacuum configuration.
The Chan-Paton group of the $\overline{{\rm D}9}$'s, say, is
\begin{equation}
SO(4) = SU(2) \times SU(2).
\end{equation}
The idea is to introduce an instanton configuration in one of
these $SU(2)$'s and accompany it by an appropriate tachyon
configuration rather like the one in the previous construction. In
this way one makes the D$5$-brane.  Moreover, the $SU(2)$, which
is not used in the instanton construction, survives as the gauge
symmetry of the D$5$-brane.  The $SO(32)$ gauge symmetry carried
by the D$9$-branes that are inert in this construction survives as
a global symmetry of the D$5$-brane world-volume theory.  (This is
also true for the D-string in the previous construction.)

\subsection{Non-BPS D$0$-brane in type I}

The perturbative $SO(32)$ heterotic string spectrum has a
spin$(32)/{\bf Z}_2$ spinor representation at the first excited
level. Even through it is not BPS, and belongs to a long
representation of the supersymmetry algebra, it is nonetheless
absolutely stable as a consequence of charge conservation.  In
heterotic string units, its mass is
\begin{equation}
M_{\rm spinor} = \sqrt{2} f(g_h),
\end{equation}
where $f(g_h) = 1 + O (g_h^2)$.  Sen~\cite{sen141} has posed the
question ``what happens to this state as $g_h \rightarrow
\infty$?''  For a BPS state, the answer would be trivial, as the
charge would control the mass for all values of the coupling.
However, this state is not BPS, so the answer is not at all
obvious.  However, one can hope to answer the question because
large $g_h$ corresponds to small $g = 1/g_h$ type I coupling. Thus
what one needs to do is to identify a plausible candidate for the
lightest gauge group spinor in the type I setting and determine
its mass to leading order in $g$.

Ordinarily, the type IIB theory has no stable D$0$-branes, though
in Section 3.3 we discussed how to construct an unstable
D$0$-brane.  However, in the presence of an orientifold (or
orbifold) plane they can exist as stable particles embedded in the
orientifold (or orbifold) plane.  One viewpoint is that the
orientifold projection eliminates the tachyonic modes from the
world-volume of the unstable type II D-brane.  Sen has studied
these possibilities for both O5-planes and ${\rm O}9$-planes. Here
we will only consider the latter case, which is realized by the
uncompactified type I theory.

We will construct a type I candidate for the desired gauge group
spinor as a D$0$-brane.  The procedure will be based on a setting
like that of the preceding sections.  Namely, following Sen, we
will construct a D$1 + \overline{{\rm D}1}$ system with a field
configuration that describes a stable soliton.

As we have already remarked, the gauge group on the world volume
of $N$ coincident type I D-strings is $O(N)$.  In particular, for
the case of a single D-string, it is $O(1) = {\bf Z}_2$. This
${\bf Z}_2$ is the subgroup of the $U(1)$ of a type IIB D-string
that survives the orientifold projection.  Being discrete, this
gauge group has no associated field, but it does have physical
consequences. The point is that the gauge group determines the
possible Wilson lines when the string is wrapped on a circular
compact dimension. Thus the possible Wilson lines for a wrapped
D-string are $W = \pm 1$.  The D-string has 32 left-moving fermion
fields $\lambda^A$ on its world sheet, which arise as the zero
modes of D$1-{\rm D}9$ open strings.  (This is also known from the
identification of the D-string with the heterotic string.)  The
possible Wilson lines (or holonomies) $W=\pm 1$ correspond to
whether the $\lambda^A$ fields of the wrapped D-string are
periodic or antiperiodic
\begin{equation}
\lambda^A (x + 2 \pi R) = W\lambda^A (x).
\end{equation}
When $W = 1$, the $\lambda^A$'s have zero modes in their Fourier
expansions, which obey a Clifford algebra.  Therefore, the quantum
states of a wrapped D-string with $W=1$ belong to the spinor
conjugacy class of Spin$(32)/{\bf Z}_2$. Similarly, the states of
a wrapped D-string with $W = - 1$ belong to the adjoint (or
singlet) conjugacy class.

Now consider a D$1 + \overline{{\rm D}1}$ pair wrapped on the
circle and coincident in the other dimensions.  To get an overall
gauge group spinor one of the strings should have $W = 1$ and the
other should have $W = -1$.  In this case it is clear (by charge
conservation) that complete annihilation of the strings is not
possible.  The wrapped D$1 + \overline{{\rm D}1}$ system contains
a real tachyon field $T$ with a potential $V(T)$.  $T$ is odd
under the action of either of the ${\bf Z}_2$ gauge groups, and
therefore gauge invariance requires that $V(T) = V(-T)$.  By the
same reasoning as before, $V$ should have an isolated minimum at
$|T| = T_0$ with $V(T_0) + 2T_{{\rm D}1} = 0$.

When both Wilson lines are the same, so that their product is $+1$
and the system can be a gauge group singlet, $T(x)$ is periodic
and it is possible to choose $T(x) = T_0$.  This would then be
equivalent to pure vacuum.  However, when the product of the
Wilson lines is $-1$, $T(x)$ is forced to be antiperiodic
\begin{equation}
T (x + 2\pi R) = - T(x),
\end{equation}
so that $T(x) = T_0$ is no longer a possibility.  Indeed, in this case the
tachyon field has a Fourier series expansion of the form
\begin{equation}
T(x,t) = \sum_n T_{n + 1/2} (t) e^{i(n + 1/2)x/{R}} .
\end{equation}
This tells us that the KK mass of the $n$th mode is
\begin{equation}
M_n^2 = \frac{(n+1/2)^2}{R^2} - \frac{1}{2}.
\end{equation}
This shows that there is actually no instability for $R < R_c$, where
\begin{equation}
R_c = 1/\sqrt{2}.
\end{equation}
In this case the two-particle system is the ground state.  There
is no lighter bound state with the same quantum numbers.  On the
other hand, when $R > R_c$, $T_{\pm 1/2}$ are tachyonic signaling
an instability.  In this case there is a bound state --- the
D$0$-brane --- that has the same quantum numbers and lower mass.

The mass of the D$0$-brane can be estimated for small $g$ by
considering the transition point $R = R_c$.  At this point the
mass of two wrapped strings and the mass of the D$0$-brane should
be degenerate.  The two wrapped strings have a mass that is
approximately given by
\begin{equation}
M = 2 \cdot 2\pi R_c \cdot T_{{\rm D}{1}} = \sqrt{2}/g.
\end{equation}
Sen argues that this result is exact to leading order in small
$g$, and that any (possibly $R$-dependent) corrections are of
$O(1)$.  As an example of such a correction consider a localized
D$0$-brane on the circle.  It experiences gravitational
interactions with its images, which are of order $G_N M^2$.
However, $M \sim 1/g$ and $G_N \sim g^2$, so this is $O(1)$. Other
effects, due to Wilson lines or KK momenta, are also $O(1)$.

We have now determined the answer to Sen's question.  The mass of
the lightest gauge group spinor is a non-BPS D$0$-brane of the
type I theory whose mass (in type I units) is $\sqrt{2}/g + O(1)$.
To compare to the perturbative heterotic result, we must still
convert to heterotic string units.  Then we conclude that
\begin{equation}
f(g_h) \sim \sqrt{g_h} \quad {\rm for ~large} ~g_h.
\end{equation}

To complete the story let us describe the D$0$-brane soliton of
the D$1 + \overline{{\rm D}1}$ system in the decompactification
limit. In this case we have a localized kink, like the ones we
described in Section 3.3.  In the present case $T(x)$ is real and
the vacuum manifold $|T| = T_0$ consists of two distinct points.
Thus the kink configuration is topologically nontrivial.  It
follows that the resulting D$0$-brane carries a conserved ${\bf
Z}_2$ charge. This means that whereas they are individually
stable, they are TCP self-conjugate and can annihilate in pairs.
This meshes nicely with the fact that they are gauge group
spinors.

\section{K-Theory Classification of D-Branes}

\subsection{Motivation and background}

Since D-branes carry conserved charges that are sources for RR
gauge fields, which are differential forms, one might suppose that
the charges could be identified with cohomology classes.  This is
roughly, but not precisely, correct.  Moore and Minasian noted
that the more precise mathematical identification of D-brane
charges is with elements of K-theory groups~\cite{moore230}.  This
was explained in detail in a subsequent paper by
Witten~\cite{witten188}. Here, we will summarize some of the basic
ideas, though we will not go deeply into the mathematics.

Witten's explanation begins with the observation that the pattern
of BPS D-branes is reminiscent of the Bott periodicity rules for
homotopy groups.  For example,
\begin{equation}
\pi_i (U(N)) = \pi_i (U(N + 2)),
\end{equation}
when $N$ is sufficiently large (the ``stable regime'').  This
pattern is reminiscent of the fact that BPS D-branes occur for
even values of $p$ in the IIA theory and odd values of $p$ in the
IIB theory, and that the gauge group of $N$ coincident type II
D-branes is $U(N)$.

The Bott periodicity rule
\begin{equation}
\pi_i (O(N)) = \pi_{i\pm 4} (Sp(N)),
\end{equation}
for sufficiently large $N$, is reminiscent of the pattern of BPS
D-branes in the type I theory.  In type I, $N$ coincident
D$1$-branes or D$9$-branes have an $O(N)$ gauge group (roughly),
whereas $N$ coincident D$5$-branes have an $Sp(N)$ gauge group.

The challenge is to find the right mathematical framework to
explain these ``coincidences'' and to account for the non-BPS
D-branes, as well.  K-theory turns out to be the key.  It provides
a classification of pairs of vector bundles with equivalence
relations that mesh nicely with what we have learned from Sen's
studies.

Let us first discuss the relevance of homotopy groups.  The
world-volume theories are gauge theories, which (as we have seen)
can support solitons. These solitons tend to be stable when there
is a topologically nontrivial lump in codimension $n$.  The field
configuration represents a nontrivial element of $\pi_{n-1} (G)$
that is identified as the conserved RR charge.  Thus, for example,
in the case of the non-BPS D$0$-brane we used the fact that $\pi_0
({\bf Z}_2) = {\bf Z}_2$ to describe a codimension one lump on a
D$1 + \overline{{\rm D}1}$ system.  Similarly, we used $\pi_3
(SU(2)) = {\bf Z}$ to construct the D-string in a D$5 +
\overline{{\rm D}5}$ system. In the type II theories, the vortex
solutions utilized the fact that $\pi_1 (U(1)) = {\bf Z}$.

Instead of the constructions listed above, we could have
constructed all of these objects as suitable bundles on space-time
filling nine-branes.  From that point of view, we would understand
the existence of the type I D-string as a consequence of
\begin{equation}
\pi_7 (O(32)) = {\bf Z},
\end{equation}
and the non-BPS D$0$-brane as a consequence of
\begin{equation}
\pi_8 (O(32)) = {\bf Z}_2.
\end{equation}

Not everything is group theory and topology, however.  Dynamics
also matters. When the codimension $n$ is greater than four, as in
these two cases, the lump cannot be described as an extremum of
the low-energy effective action
\begin{equation}
S_{\rm eff} = \int d^n x\, {\rm tr} (F_{ij} F^{ij}).
\end{equation}
To see this, suppose $A_i(x)$ is a field configuration of the
desired topology. (More precisely one defines a bundle by a
collection of $A_i$'s on open sets together with suitable
transition functions.)  Now let's rescale the gauge field as
follows
\begin{equation}
A_i(x) \rightarrow \lambda A_i (\lambda x),
\end{equation}
which sends $F_{ij} (x) \rightarrow \lambda^2 F_{ij} (\lambda x)$.
Thus, one finds that $S_{\rm eff} \rightarrow \lambda^{4-n} S_{\rm
eff}$. Therefore, for $n > 4$ the action (or energy) is reduced by
shrinking the soliton.  Presumably it shrinks to string scale
where it is stabilized by the competing effects of
higher-dimension corrections to $S_{\rm eff}$.  Thus, the string
scale should characterize the thickness of the D-string or the
D$0$-brane.

Suppose, on the other hand that the soliton has co-dimension $n < 4$.  In this
case the same scaling argument would suggest that the soliton wants to spread
out and become more diffuse.  This doesn't always happen, because there can be
other effects that stabilize the soliton, but it is an issue to be considered.

\subsection{Type II D-branes}

Consider now a collection of coincident type II D-branes --- $N$
D$p$-branes and $N'~\overline{{\rm D}p}$-branes.  As we discussed
in Section 3.1, the important world-volume fields can be combined
in a superconnection
\begin{equation}
{\mathcal A} = \left(\begin{array}{cc} A & T\\ \bar T &
A'\end{array} \right),
\end{equation}
where $A$ is a connection on a $U(N)$ vector bundle $E$, $A'$ is a
connection on a $U(N')$ vector bundle $E'$, and $T$ is a section
of $E^* \otimes E'$.  The ($p+1)$-dimensional world-volume of the
branes, $X$, is the base of $E$ and $E'$.

As we have discussed at some length in Section 3, if $E$ and $E'$ are
topologically equivalent $(E \sim E')$ complete annihilation should be
possible.  This requires $N = N'$ and a minimum of the tachyon potential $T =
T_0$, where
\begin{equation}
V(T_0) + 2NT_{{\rm D}p} = 0.
\end{equation}

As a specific example, consider the case $p = 9$ in the type IIB
theory. Consistency of the quantum theory (tadpole cancellation)
requires that the total RR 9-brane charge should vanish, and thus
$N = N'$.  So we have an equal number of D$9$-branes and
$\overline{{\rm D}9}$-branes filling the 10d spacetime $X$.
Associated to this we have a pair of vector bundles $(E, E')$,
where $E$ and $E'$ are rank $N$ complex vector bundles.  We now
want to define equivalence of pairs $(E, E')$ and $(F, F')$
whenever the associated 9-brane systems can be related by
brane-antibrane annihilation and creation.  In particular, $E \sim
E'$ corresponds to pure vacuum, and therefore
\begin{equation}
(E, E') \sim 0 \Leftrightarrow E \sim E'.
\end{equation}
If we add more D$9$-branes and $\overline{{\rm D}9}$-branes with
identical vector bundles $H$, this should not give anything new,
since they are allowed to annihilate.  This means that
\begin{equation}
(E \oplus H, E' \oplus H) \sim (E, E').
\end{equation}
In this way we form equivalence classes of pairs of bundles. These
classes form an abelian group.  For example, $(E', E)$ belongs to
the inverse class of the class containing $(E, E')$. If $N$ and
$N'$ are unrestricted, the group is called $K(X)$. However, the
group that we have constructed above is the subgroup of $K(X)$
defined by requiring $N = N'$.  This subgroup is called
$\tilde{K}(X)$.  Thus type IIB D-brane charges should be
classified by elements of $\tilde{K}(X)$.  Let's examine whether
this works.

The formalism is quite general, but to begin we will only consider
the relatively simple case of D$p$-branes that are hyperplanes in
flat ${\bf R}^{10}$. For this purpose it is natural to decompose
the space into tangential and normal directions
\begin{equation}
{\bf R}^{10} = {\bf R}^{p+1} \times {\bf R}^{9-p},
\end{equation}
and consider bundles that are independent of the tangential ${\bf
R}^{p+1}$ coordinates.  If the fields fall sufficiently at
infinity, so that the energy is normalizable, then we can add the
point at infinity thereby compactifying the normal space so that
it becomes topologically a sphere $S^{9-p}$.  Then the relevant
base space for the D$p$-brane bundles in $X = S^{9-p}$.  We can
now invoke the mathematical results:
\begin{equation}
\tilde{K} (S^{9-p}) = \left\{\begin{array}{cc} {\bf Z} & p = {\rm
odd}\\ 0 & p = {\rm even} \end{array}.\right.
\end{equation}
This precisely accounts for the RR charge of all the stable (BPS)
D$p$-branes of the type IIB theory on ${\bf R}^{10}$.  The
relation to homotopy is
\begin{equation}
\tilde{K} (S^n) = \pi_{n-1} (U(N)) \quad ({\rm large}~N).
\end{equation}
Note that $N$ does not appear on the left-hand side.  K-theory
groups are automatically in the stable regime.  It should also be
noted that the unstable type IIB D-branes, which we discussed in
Section 3.3, carry no conserved charges, and they do not show up
in this classification.

Suppose now that some dimensions form a compact manifold $Q$ of
dimension $q$, so that the total spacetime is ${\bf R}^{10-q}
\times Q$. Then the construction of a D$p$-brane requires
compactifying the normal space ${\bf R}^{9-p-q} \times Q$ to give
$S^{9-p-q} \times Q$. This involves adjoining a copy of $Q$ at
infinity.  In this case the appropriate mathematical objects to
classify D-brane charges are relative K-theory groups $K(S^{9-p-q}
\times Q, Q)$~\cite{bergman160}. In particular, if $Q = S^1$, we
have $K(S^{8-p} \times S^1, S^1)$. Mathematically, it is known
that this relative K-theory group can be decomposed into two
pieces
\begin{equation}
K(X \times S^1, S^1) = K^{-1} (X) \oplus \tilde{K} (X).
\end{equation}
The physical interpretation of this formula is very nice.
$\tilde{K} (S^{8-p})$ classifies the type IIB D-branes that are
wrapped on the circle, whereas
\begin{equation}
K^{-1} (S^{8-p}) \cong \tilde{K} (S^{9-p}),
\end{equation}
classifies unwrapped D-branes.  So, altogether, in nine dimensions
there are additive D-brane charges for all $p < 8$.

The type IIA case is somewhat more subtle, since the spacetime
filling D$9$-branes are unstable in this case.  Also, they are TCP
self-conjugate.  The right K-theory group was conjectured by
Witten, and subsequently explained by Ho\v{r}ava~\cite{horava135}.
The answer is $K^{-1} (X)$, which (as we have already indicated)
has
\begin{equation}
K^{-1} (S^{9-p}) = \left\{\begin{array}{cl} {\bf Z} & {\rm for} ~p
= {\rm even}\\ 0 & {\rm for} ~p = {\rm odd}. \end{array}\right.
\end{equation}
This accounts for all the stable type IIA D$p$-branes embedded in
${\bf R}^{10}$. The relation to homotopy in this case is
\begin{equation}
K^{-1} (S^n) = \pi_{n-1} \left(\frac{U(2N)}{U(N) \times
U(N)}\right) \quad ({\rm large}~ N).
\end{equation}
Let me refer you to Ho\v{r}ava's paper for an explanation of these
facts.

Compactifying the type IIA theory on a circle gives the relative K-theory group
\begin{equation}
K^{-1} (X \times S^1, S^1) = \tilde{K} (X) \oplus K^{-1} (X).
\end{equation}
This time $K^{-1}(X)$ describes wrapped D-branes and $\tilde{K}
(X)$ describes unwrapped ones.  This result matches the type IIB
result in exactly the way required by T duality (wrapped
$\leftrightarrow$ unwrapped)~\cite{bergman160}.

\subsection{Type I D-branes}

Type I D-branes charges can also be classified using
K-theory~\cite{witten188}. In this case we should consider $N +
32$ D$9$-branes with an $O(N + 32)$ vector bundle $E$ and $N$
$\overline{{\rm D}9}$-branes with an $O(N)$ vector bundle $E'$.
We define equivalence classes as before
\begin{equation}
(E,E') \sim (E \oplus H, E' \oplus H),
\end{equation}
where $H$ is an arbitrary $SO(k)$ vector bundle on $X$.  These
equivalence classes define the elements of a K-theory group.  If
rank $E$ $-$ rank $E'$ were unrestricted the group would be
$KO(X)$. One can define a subgroup of $KO(X)$, called
$\widetilde{KO} (X)$, by requiring rank $E =$ rank $E'$.  However,
this is not quite what we want.  The type I theory requires rank
$E =$ rank $E' + 32$.  This is a coset isomorphic to
$\widetilde{KO} (X)$, so as far as K-theory is concerned the fact
that the type I theory has 32 extra D$9$-branes is irrelevant.
However, later we will show that it is quite relevant to some of
the physics.  So we already see that K-theory is not the whole
story.

In any case, by the same reasoning as before, the conserved
charges of type I D-branes in ${\bf R}^{10}$ should be determined
by the groups $\widetilde{KO}(S^{9-p})$.  The connection to
homotopy in this case is
\begin{equation}
\widetilde{KO} (S^n) = \pi_{n-1} (O(N)) \quad ({\rm large}~N).
\end{equation}
The results are as follows:
\begin{itemize}
\item $\widetilde{KO} (S^{9-p}) = {\bf Z}$ \quad for $p =  1, 5, 9$.
\end{itemize}
These are the three kinds of BPS D-branes which carry additive
conserved charges.
\begin{itemize}
\item $\widetilde{KO} (S^{9-p}) = {\bf Z}_2$ \quad for $p = - 1, 0, 7, 8$.
\end{itemize}
These are candidates for non-BPS D-branes with a multiplicative
$({\bf Z}_2)$ conserved charges.  The $p=0$ case corresponds to
the non-BPS D$0$-brane discussed in Section 3.5.  The $p = - 1$
case is a type I D-instanton.  The $p = 7$ and $p = 8$ cases will
be discussed in Section 6.
\begin{itemize}
\item $\widetilde{KO} (S^{9-p}) = 0$ \quad for $p = 2, 3, 4, 6$.
\end{itemize}
There are no conserved D-brane charges in these cases.

Let us now consider compactifying the type I theory on a circle,
so that the spacetime is ${\bf R}^9 \times S^1$.  In this case we
want to understand the classification of D-brane charges in nine
dimensions and the T-dual description in terms of type I$^\prime$
theory.  As in the type II cases, the K-theory description of
D-brane charges of the compactified theory is given by the
relative K-theory group $KO(X \times S^1, S^1)$.  As in the type
II case, the mathematical identity
\begin{equation}
KO(X \times S^1, S^1) = \widetilde{KO}^{-1} (X) \oplus
\widetilde{KO} (X),
\end{equation}
agrees with our physical expectations.  Specifically,
$\widetilde{KO} (S^{8-p})$ classifies the wrapped D-branes and
$\widetilde{KO}^{-1} (S^{8-p})$ describes unwrapped D-branes.

There is a slightly subtle point.  The K-theory group elements
correspond to conserved charges and not to stable D-branes.  In
Section 3.5 we saw that the type I non-BPS D$0$-brane decays into
two particles, which can be described as a wrapped D$1 +
\overline{{\rm D}1}$ system when the compactification radius $R <
1/\sqrt{2}$.  The wrapped D-string that has the trivial Wilson
line $(W = 1)$ is a gauge group spinor, and it carries the charge
in question.  The K-theory classification of charges does not
distinguish the cases $R < 1/\sqrt{2}$ and $R > 1/\sqrt{2}$. While
it classifies charges, it doesn't identify which object is the
ground state with that charge.

Let us now examine the situation from the T-dual type I$^\prime$
perspective~\cite{bergman160}. In particular, we would like to
achieve a qualitative understanding of the transition at $R =
1/\sqrt{2}$ $ (R' = \sqrt{2})$. To transcribe the picture to the
type I$^\prime$ perspective recall that under T duality wrapped
D-branes map to unwrapped D-branes and vice versa. Recall, too,
that the position of an unwrapped D-brane is encoded in a Wilson
line of the corresponding wrapped D-brane.  Thus, a non-BPS
D$0$-brane of type I localized on the circle (for $R >
1/\sqrt{2}$) should correspond to a non-BPS D$1$-brane of type
I$^\prime$ stretched across the interval from $X' = 0$ to $X' =
\pi R'$.  A Wilson line of the $U(1)$ gauge group on this string
should encode the position of the D$0$-brane.

If $R < 1/\sqrt{2}$ (and $R' > \sqrt{2}$), on the other hand, then
in the type I picture one has a wrapped D$1 + \overline{{\rm D}1}$
pair. One of the strings has $W = + 1$ and the other one has $W =
- 1$. These strings correspond to D$0$-branes in the dual type
I$^\prime$ picture, and the Wilson lines tell us that those
D$0$-branes are stuck to the orientifold planes.  Thus, to be
specific, there is a $\overline{{\rm D}0}$-brane stuck to the
${\rm O}8$-plane at $X' = 0$ and a D$0$-brane stuck to the ${\rm
O}8$-plane at $X' = \pi R'$.  An interesting question is how such
a configuration morphs into a string stretched across the interval
as $R'$ is decreased through the value $\sqrt{2}$.  I have a
conjecture for the answer to this question, which goes beyond the
perturbative framework in which we have been working.  It is
reminiscent of similar phenomena found in other contexts, and is
the only smooth way that I can imagine the transition taking
place.  The idea is that as $R'$ approaches $\sqrt{2}$ from above,
the orientifold planes develop spikes so that the D$0$ and
$\overline{{\rm D}0}$ approach one another.  Then in the limit $R'
\rightarrow \sqrt{2}$ they touch and annihilate, leaving a
connecting tube between the ${\rm O}8$-planes, which in the
perturbative limit is identified as a string.  This is somewhat
analogous to the joining/breaking transition of QCD flux tubes
with the annihilation/creation of $q\bar q$ pairs.  It is also
reminiscent of a description of fundamental strings ending on
D-branes in terms of a soliton field configuration on the D-brane
world-volume~\cite{callan147,gibbons027}.  However, it differs
from these examples in a rather peculiar way.  In this case, the
inside of the tube, which is identified as a stretched non-BPS
D-string, is not even part of the spacetime!

We have just seen that there can be D$0$-branes stuck to ${\rm
O}8$-planes in the type I$^\prime$ theory.  This is T-dual to the
fact that a wrapped type I D-string can have Wilson line $W = \pm
1$, with the two possibilities corresponding to the two
orientifold planes.  Incidentally, this implies a certain
asymmetry between the orientifold planes, since $W = 1$ gives a
gauge group spinor and $W = - 1$ does not.

The bulk of the type I$^\prime$ spacetime is indistinguishable
from type IIA spacetime, at least in the region where half of the
D$8$-branes are to the left and half are to the right. (In other
regions one has a ``massive'' type IIA spacetime of the type
discovered by Romans~\cite{romans86}.) We know that the type IIA
theory has a D$0$-brane, so this suggests that the type I$^\prime$
theory should also admit this possibility. The way this works is
that a pair of stuck D$0$-branes on an orientifold plane can pair
up and move into the bulk, where the composite object is
identified as a single bulk D$0$-brane.  It is instructive to
understand the T-dual type I description of this process.

Consider a pair of wrapped type I D-strings.  The world-volume
gauge group is $O(2)$, which is nonabelian.  This system
corresponds to D$0$-branes in the type I$^\prime$ description,
with positions controlled by the choice of $O(2)$ Wilson line.
Inequivalent choices of Wilson line are classified by conjugacy
classes of the group.  This group has two types of conjugacy
classes.  A class of the first type describes a rotation by
$\theta$ or $2\pi - \theta$, which are equivalent, where $0 \leq
\theta \leq \pi$. This class corresponds to a composite bulk
D$0$-brane located at $X' = \theta R'$ and its mirror image at $X'
= (2\pi - \theta) R'$.  The remaining conjugacy class contains all
reflection elements of $O(2)$.  A representative of this class is
$\left(\begin{array}{cc} 1 & 0\\ 0 & -1 \end{array} \right)$. Thus
this class corresponds to having one stuck D$0$-brane on each
${\rm O}8$-plane.  Note that the two-body system described by this
system is a gauge group spinor, whereas a one-body system
described by a $\theta$ class is not a gauge group spinor.  I have
reviewed some additional related issues
elsewhere~\cite{schwarz061}.

\section{Moduli Spaces of Theories with 16 Supercharges}

Consistent vacua of string theories typically form spaces, called moduli
spaces, which are parameterized by the vacuum values of massless scalar fields
with flat potentials.  For theories with 16 unbroken supersymmetries (16
conserved supercharges) such as the type I and heterotic theories, the vacuum
moduli spaces are always of the Narain type
\begin{equation}
{\mathcal M}_{m,n} = \Gamma_{m,n}({\bf Z}) \backslash
SO(m,n)/SO(m) \times SO(n).
\end{equation}
Here $\Gamma_{m,n}({\bf Z})$ is the standard infinite discrete
duality group given by integral $SO(m,n)$ matrices.  Equivalently,
it can be described as the subgroup of $SO(m,n)$ that preserves an
appropriate lattice of signature $(m,n)$.  When $d\geq 4$, $n = 10
- d$ $U(1)$ gauge fields belong to the supergravity multiplet. The
remainder of the gauge fields belong to vector supermultiplets and
form a gauge group of rank $m$.  At generic points in the moduli
space this group is $[U(1)]^m$ but there is nonabelian symmetry
enhancement on various subsurfaces of the moduli space. The case
$d = 10$ is special in that the moduli space consists of just two
points --- corresponding to the $E_8 \times E_8$ and
Spin(32)/${\bf Z}_2$ theories.

\subsection{Three components in $ d= 9$}

In $ d= 9$ the moduli space of consistent vacua with 16 unbroken
supersymmetries turns out to have three disconnected components.
They correspond to the Narain moduli spaces ${\mathcal M}_{17,1}$,
${\mathcal M}_{9,1}$, and ${\mathcal M}_{1,1}$.  Since $m - n$ is
a multiple of $8$ in each case, they correspond to even self-dual
lattices, which is the key to establishing modular invariance of
the corresponding one-loop amplitudes. These three cases turn out
to have amusing geometric descriptions in terms of 11-dimensional
M theory.  They correspond to compactifications ${\bf R}^9 \times
K$, where $K$ is a cylinder for $m = 17$, a M\"obius strip for $m
= 9$, and a Klein bottle for $m = 1$~\cite{dabholkar178}.  We
observe that each boundary component of $K$ contributes 8 units of
rank to the gauge groups. The M theory viewpoint will not be
pursued further in these lectures, because our focus is on
perturbative superstring descriptions, and the M theory picture is
nonperturbative.

The case $m = 17$ also corresponds to compactification of the type
I theory on a circle.  We have discussed this system, including
its D-brane spectrum and the dual type I$'$ description in the
preceding section, so we will not say more about it here.  The
case $m = 1$ has a perturbative superstring description in $d =
9$, which will be discussed in the next subsection.  The case $m =
9$ does not have a perturbative superstring description in $d=9$
(even through it does exist).  However, a further compactification
to a ${\mathcal M}_{10,2}$ moduli space in $d=8$ makes it amenable
to perturbative superstring analysis.  We will say a little about
this case in Section 5.3.

One of our main concerns is to identify the spectrum of D-branes
and D-brane charges in each case.  The BPS D-branes are always
easy to identify, and are the same for all the components of the
moduli spaces (that have a perturbative description).  The point
is that the $C_{\mu\nu}$ RR gauge field of the type I theory in $d
= 10$ gives rise to the relevant RR gauge fields in the lower
dimensions.  Knowing these gauge fields one can immediately read
off the corresponding spectrum of BPS D-branes.  For example, in
$d = 9$ one has $C_{\mu\nu}$ and $C_{\mu 9} \equiv C_\mu$.  As a
result there are BPS D-branes in $d = 9$ for $p = 0, 1, 4, 5$ for
both the $m = 17$ and the $m = 1$ branches of the moduli space.
There are corresponding stable $p$-branes in the $m = 9$ case, as
well.  However, in that case there is no perturbative limit, so it
is not very meaningful to call them D-branes. In addition, each
case requires a certain number of BPS D$8$-branes for quantum
consistency.

\subsection{The type $\tilde{{\rm I}}$ theory}

There is a perturbative superstring description of the ${\mathcal
M}_{1,1}$ component of the $d = 9$ moduli space, which has been
called the type $\tilde{{\rm I}}$ theory~\cite{gimon038}.  It is
formulated as a type IIB orientifold that differs somewhat from
the usual type I construction.  One starts with the IIB theory on
${\bf R}^9 \times S^1$, and then mods out by the ${\bf Z}_2$
symmetry
\begin{equation}
\tilde{\Omega} = \Omega \cdot S_{1/2}.
\end{equation}
$\Omega$ is the usual world-sheet orientation reversal.  But now
it is accompanied by the operation $S_{1/2}$, which is translation
half way around the circle.  Because of this half shift,
$\tilde{\Omega}$ has no fixed points, and as a result the type
$\tilde{{\rm I}}$ theory has no orientifold planes. Therefore, no
spacetime-filling D-branes should be added to cancel the RR charge
of O-planes.  It follows that there is no gauge group, analogous
to $SO(32)$, associated to such D-branes.  Indeed the only gauge
symmetry of the 9d theory is $[U(1)]^2$, where these fields arise
from the components $g_{\mu 9}$ and $C_{\mu 9}$ of the 10d metric
and RR gauge fields.  This is what one expects for ${\mathcal
M}_{1,1}$.  The T-dual description of the type $\tilde{{\rm I}}$
theory is called type $\tilde{{\rm I}}'$.  It consists of an
interval $0 \leq X' \leq \pi R'$ (for the same reason as type
I$^\prime$).  The interval is bounded by orientifold planes.
However, unlike the type I$^\prime$ case, now the orientifold
planes carry opposite RR charge, and could be denoted ${\rm O}8^+$
and ${\rm O}8^-$.  As a result, the total RR charge vanishes and
no parallel D$8$-branes should be added.

We have already argued that the type $\tilde{{\rm I}}$ theory has
the same spectrum of BPS D-branes as the type I theory
compactified on a circle.  We will now examine the spectrum of
non-BPS D-branes and show that it differs from the spectrum of the
compactified type I theory. Bergman, Gimon, and Ho\v{r}ava
examined this problem~\cite{bergman160} and showed that the
relevant K-theory group is $\widetilde{KSC}(X)$ and that this
predicts non-BPS D$p$-branes carrying a ${\bf Z}_2$ charge for $p
= - 1, 3, 7$.  I refer you to their paper for the K-theory
analysis.  What I want to describe here is a physical argument
they presented in support of this conclusion.

Because we mod out by $\tilde{\Omega}$, a D$p$-brane localized on
the circle must be accompanied by an $\Omega$-reflected D$p$-brane
at the opposite location half way around the circle. Now, the
action of $\Omega$ on type IIB D-branes is as follows:
\begin{eqnarray}
\Omega: \quad {\rm D}p & \rightarrow & {\rm D}p 
\qquad p = 1,5,9\nonumber \\
\Omega: \quad {\rm D}p & \rightarrow & 
\overline{{\rm D}p} \qquad p = -1, 3, 7.
\end{eqnarray}
This is why the type I theory only has BPS D-branes for $p = 1, 5,
9$.

The configurations based on $p = 1, 5$ give expected BPS D-branes
in 9d.  However, the configurations with a $p = - 1, 3, 7$
D$p$-brane and an accompanying $\overline{{\rm D}p}$-brane
half-way around the circle are non-BPS.  But they also correspond
to stable configurations in 9d, at least when the radius of the
circle is large enough.  (Otherwise the open string connecting
them might develop a tachyonic mode.)  So the result agrees with
the K-theory prediction.

We can see that these non-BPS D$p$-branes are conserved modulo 2,
so that they carry a ${\bf Z}_2$ charge.  Imagine two of them
localized on the circle.  Then one of them (and its image) can be
slid around the circle until the D$p$-brane of the first pair
encounters the $\overline{{\rm D}p}$-brane of the second pair.  At
this point the pairs can annihilate into neutral radiation.  As an
exercise, the reader might want to describe the T-dual type
$\tilde{{\rm I}}'$ description of this annihilation process.

\subsection{Consistent vacua in eight dimensions}

The three 9d branches of the moduli space correspond to three
branches in 8d, as well.  They are ${\mathcal M}_{18, 2}$,
${\mathcal M}_{10,2}$, and ${\mathcal M}_{2,2}$.  Below 8d there
are additional branches, which we will not discuss here.  The
three branches in 8d can be constructed from a type I perspective
by compactifying on a torus, so that the spacetime is ${\bf R}^8
\times T^2$, and allowing appropriate holonomies associated to the
two cycles of the torus.  The possible choices of holonomies were
analyzed by Witten~\cite{witten028}.  His key observation was that
it is possible to have holonomies  that correspond to gauge
bundles without vector structure.

The holonomies associated to the two cycles are elements of the gauge group,
which we can write as $32 \times 32$ matrices $g_1$ and $g_2$.  In order to
describe a flat bundle, which is a necessary requirement, there are two
possibilities
\begin{equation}
[g_1, g_2] = 0 \quad {\rm or} \quad \{g_1, g_2\} = 0.
\end{equation}
The first case is the one with vector structure, and the second is
the one without vector structure.  The point is that holonomies,
when written in terms of allowed representations of the gauge
group, should commute to give a flat bundle.  However, the
nonperturbative gauge group is Spin(32)/${\bf Z}_2$, which does
not admit the 32-dimensional vector as an allowed representation.
Matrices that anticommute in the \underline{32} can correspond to
ones that commute when expressed in terms of any of the allowed
representations.

Witten analyzed the possibilities and found that there are three
inequivalent choices, one with vector structure and two without
vector structure.  The one with vector structure is the obvious
choice, which gives ${\mathcal M}_{18, 2}$.  In each case it is
illuminating to consider the T-dual description on
$\tilde{T}^2/{\bf Z}_2$.  In this description the ${\mathcal
M}_{18,2}$ branch of the moduli space corresponds to having four
${\rm O}7$-planes, each with RR charge $-4$.  Thus 16 D$7$-branes
localized on $\tilde{T}^2/{\bf Z}_2$ must be added.  This is a
straightforward generalization of the type I$^\prime$ description
in 9d. The ${\mathcal M}_{10,2}$ branch corresponds to having
three ${\rm O}7$-planes with RR charge $-4$ and one with charge
$+4$. In this case eight D$7$-branes are required to cancel the
charge. The last case, ${\mathcal M}_{2,2}$ is described by two
${\rm O}7$-planes with RR charge $-4$ and two with charge $+4$ and
no D$7$-branes. In each case, one sees that the number of
D$7$-branes agrees with what is required to explain the rank of
the gauge group. Incidentally, if there were more than two ${\rm
O}7$-planes of positive charge, one would need $\overline{{\rm
D}7}$-branes to cancel the RR charge.  Such a configuration would
not be supersymmetric.

An exercise that has not yet been carried out is to determine the
spectrum of non-BPS D-branes for the ${\mathcal M}_{10,2}$ branch
of the moduli space. This should be possible now.

\section{Non-BPS D-Branes in Type I}

\subsection{Determination of the nonperturbative gauge groups}

The gauge symmetry of the perturbative type I theory is $O(32)$.
(The reflections have no consequence, so there is no harm in
including them.) However, the non-BPS D-instanton, which showed up
in the K-theory classification in Section 4.3, arises because
$\pi_9 (SO(32)) = {\bf Z}_2$.  Witten argued that it is
responsible for breaking $O(32)$ $\rightarrow
SO(32)$~\cite{witten188}. Moreover, the non-BPS D-particle, which
we have discussed at some length, is a gauge group spinor. More
precisely, it gives states belonging to one of the two spinorial
conjugacy classes of Spin(32).

When all of these facts are taken into account, one concludes that
the nonperturbative type I theory has a different gauge group than
is visible in perturbation theory.  Specifically, it is
Spin(32)/${\bf Z}_2$.  This agrees with the gauge symmetry that is
manifest in the perturbative heterotic theory.  This agreement can
be regarded as a successful test of S duality.  Incidentally, we
also can conclude that there are no nonperturbative effects in the
heterotic description that modify the gauge group.

\subsection{Instability of D$7$ and D$8$}

The K-theory classification of type I D-brane charges, presented
in Section 4.3, suggests the existence of a non-BPS D$7$-brane and
a non-BPS D$8$-brane, each of which is supposed to carry a
conserved ${\bf Z}_2$ charge.  However, there is a tachyon in the
spectrum of D$7 - {\rm D}9$ and D$8 - {\rm D}9$ open
strings~\cite{frau123}. Therefore, the proposed D$7$-brane and
D$8$-brane should each have 32 tachyon fields in their
world-volume theory, and therefore they must be unstable. (Such
instability occurs for D$p - {\rm D}q$ open strings whenever $0 <
|p-q| < 4$.)  This instability does not, by itself, constitute a
contradiction with the K-theory analysis, which only purports to
give the types of conserved D-brane charges, and not the specific
objects which carry them. But it is a cause for concern. Clearly,
the instability occurs in these cases because their are 32
spacetime filling D$9$-branes in the type I vacuum. As we pointed
out earlier, the K-theory analysis is indifferent to their
presence. The question is to what extent the ${\bf Z}_2$ charge
survives when the D$7$ or the D$8$ dissolves in the background
D$9$-branes.

\subsection{Further analysis of the D$8$}

By Bott periodicity, one might expect the non-BPS D$8$-brane to
have features in common with the non-BPS D$0$-brane.  There is one
essential difference, however,  The D$0 - {\rm D}9$ open strings
do not have tachyonic modes whereas the D$8 - {\rm D}9$ open
strings do. We can confirm this fact by trying to emulate Sen's
construction of the D$0$-brane as a kink solution on a D$1 +
\overline{{\rm D}1}$ system~\cite{schwarz091}.

In the case of the D$1 + \overline{{\rm D}1}$ system we saw that
there is a single real tachyon $T$ and a potential $V(T)$ with
minima at $T = \pm T_0$, which can be regarded as a 0-dimensional
sphere. The vacuum manifold has two disconnected components and,
therefore, the kink solution that connects them is topologically
stable.  The closest analog we can construct eight dimensions
higher is to consider a system consisting of 33 D$9$-branes and
one $\overline{{\rm D}9}$-brane.  The world volume in this case
has 33 tachyonic modes $\vec T$ belonging to the fundamental
representation of the $O(33)$ gauge symmetry carried by the
D$9$-branes.  The potential $V(\vec T)$ must have $O(33)$ gauge
symmetry, so the minima at $|\vec T| = T_0$ describe a $S^{32}$
vacuum manifold.  This moduli space is connected, and therefore
does not support a topologically stable kink.  The situation is
very reminiscent of the unstable type II D-branes discussed in
Section 3.3.  In that case there was an $S^1$, and as a result the
kink solution had one unstable direction, so that the resulting
D-brane ended up with one tachyonic mode.  In the present case
there are 32 unstable directions so that the resulting D$8$-brane
has 32 tachyonic modes in its world volume.  Of course, these are
the same modes that one discovers by quantizing the D$8 - {\rm
D}9$ open strings.  In the case of the unstable type II D-branes,
there was no associated conserved charge, and they did not appear
in the K-theory analysis.  The type I D$8$-brane, on the other
hand, does appear in the K-theory analysis, even though it has an
analogous instability.

Witten has argued in support of the D$8$-brane as follows:  The
D-instanton implies that there are two distinct type I vacua
distinguished by the sign of the D-instanton amplitude.  This is a
${\bf Z}_2$ analog of the $\theta$ angle in QCD.  One should
expect that there is a domain wall connecting the two vacua
--- the D$8$-brane.  The sign change in the instanton amplitude means that the
D-instanton is the electromagnetic dual of the D$8$-brane.  This
has been investigated by Gukov~\cite{gukov042}.

Even though the D$8$-brane is unstable one could imagine forming
it at some moment in time and asking how the vacua on the two
sides of its are distinguished.  Recalling that the D-instanton
was responsible for breaking $O(32) \rightarrow SO(32)$, it seems
clear that these vacua should be distinguished by a gauge-group
reflection. This implies that opposite spinor conjugacy classes
would enter into the Spin(32)/${\bf Z}_2$ gauge group on the two
sides.  Of course, once the D$8$-brane decays, the vacuum becomes
uniform everywhere.

\section{Concluding Remarks}

There has been considerable progress in analyzing
non-supersymmetric D-brane configurations.  In particular, some
stable non-BPS D-branes have been identified, and they have been
studied in perturbative limits with some mathematical control.
However, without the BPS property it is not possible (at present)
to make quantitative studies away from weak coupling.

Another significant development has been the identification of
K-theory groups as the appropriate mathematical objects for
classifying D-brane charges.  In the case of BPS D-branes, these
charges are sources for RR gauge fields, whereas in non-BPS cases
they are not.  The non-BPS D-branes that we have discussed carry
conserved ${\bf Z}_2$ charges.  However, Sen has analyzed examples
in which other charge groups (such as ${\bf Z}$) also appear. One
lesson we have learned is that K-theory does not take account of
spacetime filling D-branes, such as the 32 D$9$-branes of the type
I theory.  Their presence can destabilize other D-branes in
certain cases.

\section*{Acknowledgments}

I am grateful to Oren Bergman, Ashoke Sen, and Edward Witten for
helpful discussions and suggestions. This work supported in part
by the U.S. Dept. of Energy under Grant No. DE-FG03-92-ER40701.


\end{document}